\documentclass{svproc}
\usepackage{url}

\usepackage{graphicx} 
\usepackage{natbib}

\begin{document}
%
\title{Efficient coding with chaotic neural networks:\\A journey from neuroscience to physics and back}
\titlerunning{Coding with chaotic networks}  
%
\author{Jonathan Kadmon}
%
%
%
\institute{Eddmond and Lily Center for Brain Sciences (ELSC)
\\ 
The Hebrew University, Jerusalem
\\
\email{jonathan.kadmon@mail.huji.ac.il}}

\maketitle              

\begin{abstract}
This essay, derived from a lecture at "The Physics Modeling of Thought" workshop in Berlin in winter 2023, explores the mutually beneficial relationship between theoretical neuroscience and statistical physics through the lens of efficient coding and computation in cortical circuits.  It highlights how the study of neural networks has enhanced our understanding of complex, nonequilibrium, and disordered systems, while also demonstrating how neuroscientific challenges have spurred novel developments in physics. The paper traces the evolution of ideas from seminal work on chaos in random neural networks to recent developments in efficient coding and the partial suppression of chaotic fluctuations. It emphasizes how concepts from statistical physics, such as phase transitions and critical phenomena, have been instrumental in elucidating the computational capabilities of neural networks.

By examining the interplay between order and disorder in neural computation, the essay illustrates the deep connection between theoretical neuroscience and the statistical physics of nonequilibrium systems. This synthesis underscores the ongoing importance of interdisciplinary approaches in advancing both fields, offering fresh perspectives on the fundamental principles governing information processing in biological and artificial systems. This multidisciplinary approach not only advances our understanding of neural computation and complex systems but also points toward future challenges at the intersection of neuroscience and physics.

\end{abstract}

\section*{Introduction}
\label{introduction}
Capturing the nuances and intricacies of interdisciplinary scientific research is likely as delicate and elusive as producing fine art. When one thinks of science, math, nature, and art, one of the first names that comes to mind is the contemporary American painter Mark Tensey. Tansey's art masterfully intertwines these domains, often employing visual metaphors and allegorical scenes to explore and critique complex ideas, blending rigorous conceptual thought with meticulous craftsmanship. Of his works, the painting \emph{Achilles and Tortoise (1986)} is an exquisite example of the deep interplay between applied and basic research.

In the background, we see a towering Hemlock tree. Its leaf formation attests to nature's complexity. In front of it, a group of men, we will call them the \emph{engineers}, has just launched a rocket into the sky. The projectile is soaring upwards; its contrails resemble the silhouette of the Hemlock tree, signifying our attempts to imitate nature. The rocket is short of reaching the full height of the tree, and we are reminded of Zeno's paradox of Achilles and the Tortoise---the title of this artwork.

In the front of the scene, another group of men, \emph{the scientists}, is assembled and planting a tree. Their natural approach to creating another tree sharply contrasts with the man-made rocket of the engineers. Overlooking the scientists, we notice another group, \emph{the theorists}. The distinct figure of Albert Einstein immediately discloses the group's identity. However, we focus on the other members of the scientific band. The first is Zeno himself, dressed in a black tie. The other two are Mitchel Feingelbaum and Benoit Mandelbrot---two mathematicians instrumental to the foundation of Chaos theory.

The presence of the two mathematicians hints at Tansey's view of nature, at the boundary of order and chaos. Feingelbaum studied the universality of the transition into chaos \citep{feigenbaum1978}, and Mandelbrot showed that simple rules can lead to infinite complexity, in which the separation between order and disorder is not always clear \citep{mandelbrot1967}. Hints to this complexity are found in the shape of the Hemlock and the rocket's contrails.

Tansey's painting demonstrates our fascination with order and disorder. Chaos theory has captured the minds of many men and women in art and science due to its ability to demonstrate how complexity can arise from simplicity. It challenges our intuitive understanding of cause and effect. Modern biology has always struggled to unify the simple and the complex, and neuroscience is a prime example of such tension. On the one hand, the biology and physiology of the brain are obscenely complex, with noise and disorder appearing in every scale we observe. On the other hand, our view of the brain as a computing system requires order, reliability, and preciseness.

John von Neumann, the father of the modern computer, was the first to ponder how biological systems, with their unreliable components, can execute critical computations. In this essay, I aim to address this question. My objectives are twofold: first, to describe some of our recent efforts to reconcile the tension between reliable and efficient computation and the seemingly erratic nature of the brain, and second, to demonstrate the interplay between fundamental theories in statistical physics and neuroscience research. This essay is not a comprehensive review of the field; instead, it highlights my personal experience with the mutual fertilization of ideas that constantly occurs between these two disciplines.

\section*{The case for a chaotic brain}
\label{the-case-for-a-chaotic-brain}
The brain's complexity presents an intriguing paradox when we consider the behavior of individual neurons versus their collective activity. When isolated neurons are extracted from brain tissue and subjected to constant synaptic current, their responses are remarkably orderly and predictable \citep{pfahlert2008thoracic}. Simple models can accurately forecast these individual neural activities, suggesting minimal inherent noise in the system at this level.

However, a dramatically different picture emerges when we observe these same neurons operating within the context of a large network. In this setting, neuronal responses become highly irregular, with significant variations in both their firing patterns and rates  \citep{shadlen1998variable}. This irregularity is not merely a subjective observation but can be quantified using two key statistical measures: (1) The coefficient of variation (CV), which assesses the temporal regularity of neuronal activity; and (2) The Fano Factor, which evaluates irregularity across multiple trials by measuring the variation in responses to identical repeated stimuli.

Both of these measures show substantial increases when neurons are interconnected in a network. This stark contrast suggests that irregularity is an emergent property of the network rather than an inherent characteristic of individual neurons. In essence, neurons become less reliable predictors of behavior when functioning as part of a larger system.

These observations raise two fundamental questions. First, how does a system composed of deterministic and regular components generate such a high degree of irregularity when these components are interconnected? Second, given this inherent irregularity, how does the brain manage to perform reliable computations?

A groundbreaking study by Sompolinsky, Crisanti, and Sommers addressed the first of our key questions: How can a network composed of deterministic elements exhibit unpredictable behavior  \citep{Sompolinsky1988-pm}? This seminal paper revealed that large networks of randomly connected, deterministic nonlinear neurons can enter a regime dominated by random fluctuations. Randomly connected means that the synaptic efficacy between each pair of neurons is drawn independently from a finite-variance distribution.

The work by Sompoliosnsky et al.  demonstrated that the emergent unpredictability in neural networks isn't necessarily a result of inherent randomness in individual components, but rather it may arise from the complex interactions within the system as a whole. The authors' explanation for this phenomenon invoked a concept well-established in science and mathematics: chaos. However, the form of chaos uncovered in their study differed significantly from previously examined chaotic systems. This distinction opened up new avenues for understanding complex neural dynamics and laid the foundation for an entirely new field of study.

\section*{A new kind of chaos}
\label{a-new-kind-of-chaos}
The findings of Sompolinsky et al. were groundbreaking from several perspectives. First, they provided the first mathematical explanation for the apparent paradox where the brain, composed of seemingly reliable and deterministic units, exhibits stochastic behavior. This theory has led to numerous subsequent works that have transformed the field of theoretical neuroscience. The impact of their work remains prominent in recent research, ranging from the dynamics of cortical circuits \citep{Kadmon2015-rz} to deep learning \citep{mignacco2020dynamical}. Moreover, the influence of this seminal paper extended well beyond neuroscience and machine learning. Deterministic chaos has since been studied and observed in various networks, from genetic and cellular systems  \citep{heltberg2019chaotic} to ecology \citep{roy2019numerical}.

The paper also made novel contributions to mathematics and chaos theory. It identified a bifurcation point---a dynamical transition into chaos as one parameter changes. In this case, the parameter was the gain of the single-neuron input. The gain parameter describes how strongly the input drives a single neuron to fire. Consequently, it also determined the strength of nonlinearity in the network, as the source of nonlinearity is the input-output transformation of individual neurons.  Prior to this work, transitions between orderly and chaotic dynamics were only observed in low-dimensional systems. A classic example is the logistic map  \citep{strogatz1994}, where discrete dynamics transition from a fixed point to a chaotic attractor through a series of period-doubling events known as the "route to chaos". While chaos in high dimensions had been studied before (e.g., in high-dimensional billiards modeling the interaction of many particles), these systems did not show a smooth transition from regular to chaotic activity through a dynamic bifurcation. Furthermore, billiards and particle systems are typically studied at equilibrium, whereas neural networks are inherently out of equilibrium.

The source of chaotic dynamics in random neural networks is the disorder in connectivity. This results in unpredictable and uncorrelated fluctuations in the activity of single neurons. The work by Sompolinsky et al. demonstrated, for the first time, the connection between disorder and chaos. It showed that irregularity in the connectivity patterns between neurons, termed "quenched noise" or "frozen noise", can lead to "dynamic noise" manifested through chaotic fluctuations.

Sompolinsky's work utilizes ideas and methods from spin glasses and disordered statistical systems \citep{sompolinsky1981dynamic}. One of its greatest technical contributions was connecting dynamic mean field theory (DMFT) \citep{sompolinsky1981time}, which studies the time-dependent behavior of disordered systems, with chaos theory. This connection is established through an elegant proof showing that the only stable solution for the collective dynamics exhibits chaos and a positive mean Lyapunov exponent \citep{strogatz1994}. To this day, the results and methods developed in the work on chaos in random neural networks have implications that extend further from neuroscience and network science.

\section*{Computing with chaotic neural networks}
\label{computing-with-chaotic-neural-networks}
One of the principal benefits of chaotic neural networks is their enhanced computational power, primarily attributed to the networks\textquotesingle{} sensitivity to initial conditions. This characteristic allows these networks to function as highly effective classifiers, capable of distinguishing between minute differences in input stimuli (see, e.g., \citep{bertschinger2004real,Keup2021-hn}). Thus, the chaotic networks's sensitivity facilitates superior performance in comparative tasks, outperforming traditional computational paradigms.

Another advantage of chaotic neural networks is their robustness to noise \citep{Toyoizumi2011-sz,kadmon2020predictive}. Contrary to conventional computational models susceptible to noise-induced errors, chaotic networks exhibit superior resistance to noise interference. This resilience increases their applicability in real-world scenarios characterized by inherently noisy environments.

In terms of expressivity, chaotic networks are recognized for their adeptness as pattern generators, as highlighted in the highly recognized work by Sussillo and Abbott \citep{Sussillo2009-wt}. Additionally, they contribute to increased curve length \citep{Raghu2016-uh} and generate positive entropy production \citep{Engelken2020-kr}. These characteristics enhance the networks\textquotesingle{} ability to express and represent diverse patterns and functions. The high entropy of these networks suggests a significant capacity for information encoding. This attribute can be instrumental in handling complex, high-dimensional data sets, which is necessary in the rapidly evolving field of big data.

Nonetheless, these enticing benefits come with a significant caveat: instability. The very trait that empowers chaotic networks - sensitivity to initial conditions - can also be their Achilles heel. Minor input variations or initial conditions can lead to exponentially increasing differences over time, a feature that may wreak havoc in computations, especially those requiring precision and reliability. Asking a chaotic system to convey information faithfully is analogous to asking Jackson Pollock to paint a realistic portrait. Although artistically captivating, the lack of fine-grained control and predictability is far from desirable when precision is paramount.

\section*{At the edge of chaos}
\label{at-the-edge-of-chaos}
Chaotic networks may open the door to high-capacity computing with high expressivity. However, their unpredictability is to their detriment, seemingly preventing them from living up to their potential. Conversely, perfectly ordered systems (for example, crystals) do not have enough expressivity to support useful computation. Thus, It is reasonable to assume that an optimal computing system will operate between these two extremes. Following similar arguments, previous studies have quantified a system's complexity, arguing that complexity peaks between the orderly (crystal) and disordered (gas) phases \citep{lopez1995statistical}.

\begin{figure}
  \centering
    \includegraphics[width=\linewidth, keepaspectratio]{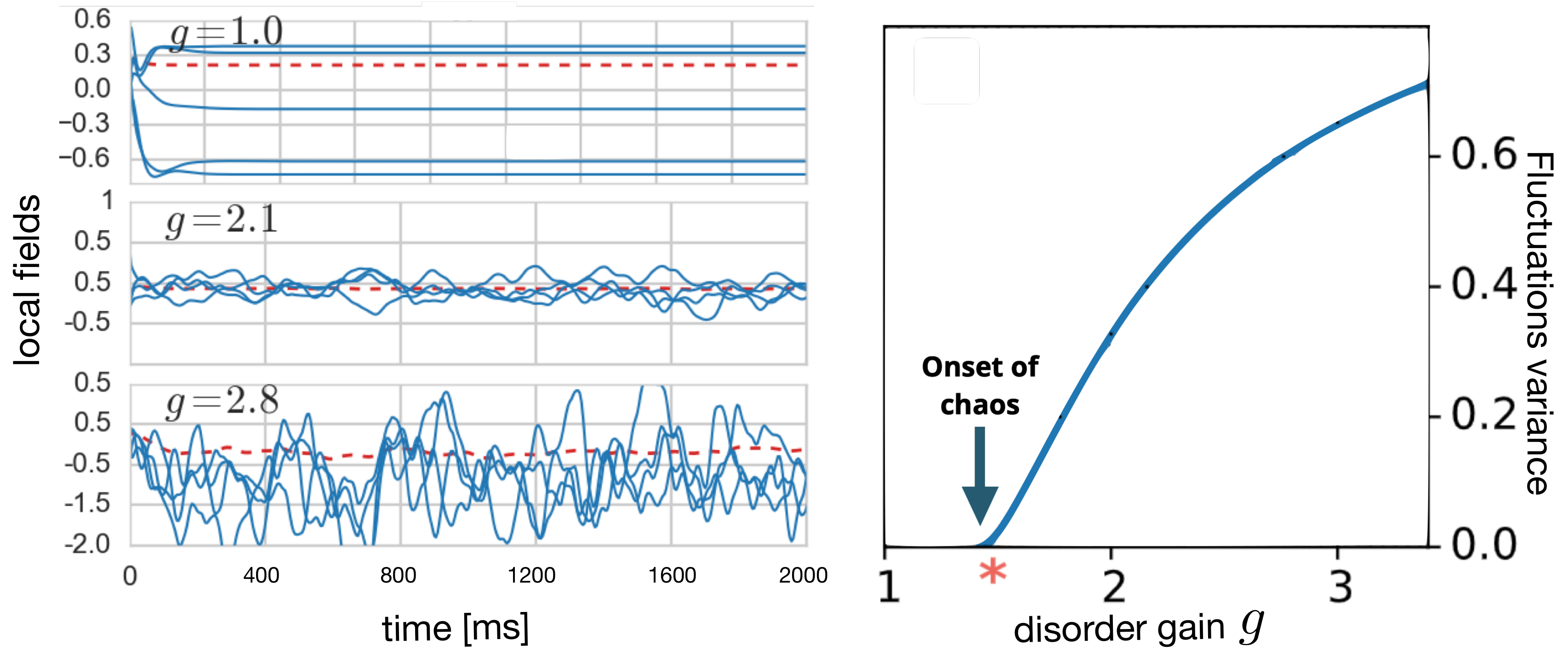}
    \caption{\textbf{Transition to chaos in large random recurrent networks.} \textbf{Left:} Time traces of neural activity (solid blue) and their mean (dahsed red)  for different levels of disorder gain ($g$). As the gain increases, the neural activity transitions from a fixed point (at low $g$) to chaotic dynamics (at higher $g$). \textbf{Right:} The order parameter, defined as the variance of fluctuation activity, plotted against the disorder gain. Below the transition to chaos (red asterisk at $g = \sqrt{2}$), the order parameter is zero, implying no random fluctuations in the network. Above the transition, the fluctuations increase with the gain. Data and plots adapted from \citep{Kadmon2015-rz}}.
    \label{fig:fig1}
  \end{figure}

Sompolinsky et al. showed that as the gain, or alternatively the disorder in the network, increases, the fixed point of the dynamics loses its stability, and the system becomes chaotic. This transition is sharp for large networks (at the thermodynamic limit). This is akin to phase transitions in physical systems, e.g., the conversion between ferromagnetic and paramagnetic states of matter as a magnetic system is cooled. Similarly, a single point demarcates the sharp boundary between the ordered and chaotic dynamics domains in large neural networks. Figure \ref{fig:fig1} shows sample trajectories of neural activity below and above the chaotic transition.
  
When the transition between chaotic and stable dynamics occurs at specific parameter values, the network can be arbitrarily close to it. This situation, where the network is at the brink of chaos, is broadly termed the ``edge of chaos.'' This term refers to carefully tuned networks close to the transition point. Preparing a system in this position requires an effort, but, as we shall argue next, this effort can be rewarding as it improves the network\textquotesingle s computational power.

In physics, phase transitions describe the transformation between different states of matter, such as from a solid to a liquid or from a liquid to a gas. These transitions occur when a system's macroscopic properties change drastically due to a variation in an external control parameter, such as temperature or pressure. A quintessential feature of phase transitions is the change in \emph{order parameters}, macroscopic quantities that vanish in one phase and are nonzero in another, signifying a fundamental change in the system's state. The order of the phase transition is defined by the discontinuity in the system's free energy: if the free energy changes are discontinuous as the system undergoes a transition, it is a first-order phase transition. If the discontinuity only happens in the first derivative of the free energy, then it is a second-order phase transition.

Second-order (or continuous) transitions are particularly notable among phase transitions. These transitions are characterized by a continuous change in the order parameter. In addition, second-order phase transitions exhibit critical phenomena, such as divergence of correlation lengths and slowing down of the system near the transition. Furthermore, second-order transitions are categorized by scaling behaviors reflecting the underlying symmetries of the transitions that transverse different system scales.

Out of equilibrium, phase transitions are not as well defined. In particular, neural networks are nonequilibrium systems that do not obey energy conservation. This is a fundamental property due to the non-symmetry in the neurons' interaction that defies Newton's third principle (unless we force the connectivity between neurons to be symmetric). From a statistical physics perspective, random neural networks do not display detailed balance \citep{vanKampen2007} and thus are not an equilibrium. In these systems, we extend the notion of phase transition to dynamical phase transitions. Dynamical phase transitions involve qualitative changes in the temporal behavior of the system as a function of parameters such as synaptic strength and input. These transitions arise from intrinsic nonlinearities and feedback mechanisms, leading to bifurcations where small parameter changes cause sudden shifts in system behavior.

Neural networks exhibit different dynamical bifurcations that separate different types of temporal behavior, including fixed points, periodic oscillations, and chaotic dynamics. While transitions between stable fixed points and orbits are well described by the theory of dynamical systems, the transition to chaos, as found by Sompolinsky et al., can only be described statistically, in the limit of very large networks (known as the \emph{thermodynamic limit}). This distinction makes the transition to chaos unique among dynamical bifurcations and more similar to a phase transition of matter.

However, since neural networks are fundamentally nonequilibrium systems, the transition to chaos cannot be clearly classified as a first or second-order transition. Nevertheless, the transition shows many characteristics of a second-order phase transition, particularly critical behavior and scaling phenomena \citep{Kadmon2015-rz}. As the disorder in the network increases, the system becomes more regular, pairwise correlations increase, and the typical temporal scales observed in the network diverge.

The critical phenomena observed near the transition to chaos potentially benefit neural computation. Near the transition, the increased regularity and prolonged temporal scales allow the system to produce coherent, longer-lasting activity patterns. The rich neural dynamics at the edge of chaos make a large reservoir of spatiotemporal patterns that can be used for downstream computation \citep{Jaeger2004-wx}. In addition, long timescales enable networks at the edge of chaos to train on tasks requiring longer temporal correlations and carry the error signal necessary for training further back in time \citep{Raghu2016-uh,Bahri2020-zx}. Other merits of networks at the edge of chaos include efficient transmission of information in a noisy environment \citep{Toyoizumi2011-sz} and discrimination between stimuli \citep{bertschinger2004real,Keup2021-hn} .

While critical phenomena can improve the usefulness of a network set near the transition to chaos, the most direct benefit is that it lies between the chaotic and the ordered phases. Near the boundary, the dynamics can easily transition between the two states with a slight change in conditions and utilize the computational benefits of chaotic dynamics while avoiding many of its drawbacks. In particular, an external input into the network can control the dynamic phase. For example, adding a static homogeneous input to a chaotic network can shift the transition point higher \citep{Kadmon2015-rz}, resulting in controllable stable dynamics. Other external inputs such as oscillation \citep{Rajan2010-vm} and even small uncorrelated noise \citep{Schuecker2016-nc} have been shown to entrain the chaotic dynamics.

\section*{Partial suppression of chaos}
\label{partial-suppression-of-chaos}
When a computational system operates at the edge of chaos, we can induce a dynamical phase transition, providing external input, thereby reducing chaotic fluctuations as needed. However, such a dynamical phase transition typically results in a global change, completely eliminating chaotic behavior and stabilizing the system. In contrast, experimental observations indicate that while the reliability of individual neurons increases during repetitive tasks, they still exhibit variability \citep{churchland2010stimulus}, suggesting that a significant stochastic or chaotic component remains present. These findings imply that the brain maintains reliable coding and random fluctuations simultaneously, and the fluctuations are removed when an efferent component reads out the signal.

To understand the suppression of chaotic activity by a readout, it is essential to consider the properties of high-dimensional chaos in large random networks. Classical chaotic systems, like the logistic map or Lorentz attractor, exhibit unpredictable fluctuations across all degrees of freedom. On the other hand, in neural networks, the dimensionality of chaotic fluctuations is relatively low, capped at roughly 10\% of the total network dimensionality, as measured by the PCA dimensions (also known as participation ratio in physics) \citep{gao2017theory}. The reduced dimensionality suggests that a significant portion of the neural dynamics space remains available for reliable coding, provided it is utilized correctly. At this point, it is crucial to recognize that neural activity is inherently nonlinear, and there is no clear separation between chaotic and nonchaotic directions. Therefore, achieving reliable coding does not simply involve isolating a subspace of neural activity; reliability is instead achieved in a statistical sense.

The partial suppression of chaotic activity directly results from the central limit theorem. Assuming that the dimensionality of the relevant signal---or the dynamics implemented by the network---is much smaller than the full neural space and that these dynamics are distributed across all neurons, the effective dimensionality of chaotic activity is significantly smaller than the full dimension of neural dynamics, yet larger than the signal dimensionality. Consequently, when averaging over \(N\) neurons with uncorrelated fluctuations, the mean fluctuation decays as $1/\sqrt{N}$. This principle also applies to random projections of neural activity. If neural activity is projected onto a random direction in the \(N\)-dimensional space, the chaotic fluctuations in that direction diminish as \(\ 1/\sqrt{N}\).

To understand how chaotic activity influences coding, it is necessary to move beyond traditional mean-field and dynamic mean-field theories, which present two significant challenges. First, chaotic fluctuations tend to vanish when \(N\) approaches infinity (the thermodynamic limit). In classical problems within statistical and condensed matter physics, the number of particles is extraordinarily large, often on the scale of Avogadro's number, allowing random fluctuations to be disregarded. However, when applying statistical physics methods to neural networks and other computing systems, we often deal with much smaller systems. In particular, brain circuits of interest often consist of just a few thousand neurons---many orders of magnitude smaller than Avogadro's number. Therefore, adapting mean-field methods to account for finite-size effects is crucial.

The second challenge arises from the nature of neural networks and computing systems, which are neither completely disordered nor entirely ordered. Furthermore, both the ordered and disordered components may play significant roles in these systems. Consequently, developing theories that can accurately describe such chimera systems is essential. One approach is to identify sufficient order parameters that faithfully capture the ordered dynamics within the system and distinguish them from the bulk of disordered degrees of freedom. Such refined theoretical frameworks enable a more precise understanding of the interplay between chaos and order in neural networks.

As a concrete example, consider the problem of efficiently encoding a signal in a chaotic neural network. In this scenario, a neural network receives an external signal, and an efferent linear readout must recover it. The challenge is that the network's activity is unreliable due to chaos or noise, leading to a corrupted signal. One approach to overcome this issue is to increase the number of neurons in the system. As argued above, the effects of the random fluctuations diminish as \(1/\sqrt{N}\). However, this solution is inefficient, requiring more neurons thereby consuming more energy and resources. Another straightforward method to improve the signal-to-noise ratio is to amplify the signal. Yet, this approach implies higher firing rates of individual neurons, which is also inefficient. Moreover, neurons can become saturated, quickly limiting the effectiveness of this method.

\begin{figure}
  \centering
    \includegraphics[width=\linewidth, keepaspectratio]{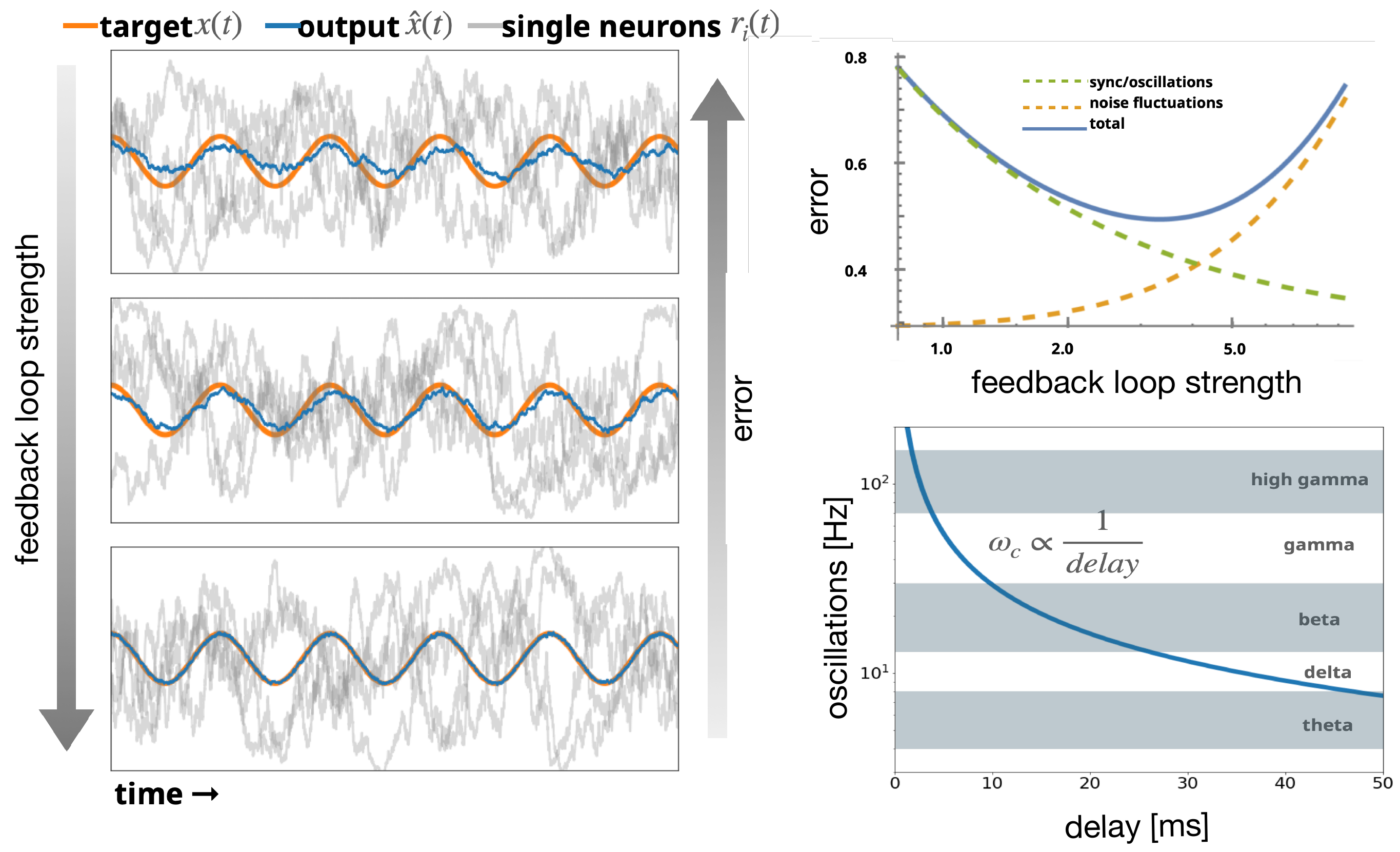}
    \caption{\textbf{Influence of Feedback Loop Strength and Delay on Neural Network Dynamics.} \textbf{Left:}  Increasing the strength of the feedback loop results in a more reliable readout, as shown by the comparison between the readout (solid blue) and the target signal (orange). Importantly, individual neurons (sampled grey traces) exhibit strong chaotic fluctuations even when the readout is reliable. \textbf{Top Right:} Amplitude of fluctuations (dashed orange) and oscillations (dashed green) as a function of the feedback loop strength (the relative strength of the structure or order in the system). The overall readout error is proportional to the sum (solid blue), which shows an optimal point. \textbf{Bottom right:} Frequency of resonant oscillations due to feedback as a function of delay time. Different delays lead to different resonant frequencies. For example, axonal delays (4-8 ms) contribute to the high-gamma band; synaptic delays (10-50 ms) contribute to beta or delta bands; and delays due to processing by other subnetworks, which can take longer, may contribute to the theta band. }
    \label{fig:fig2}
  \end{figure}

A more efficient strategy to enhance the network's performance is to utilize recurrent connectivity to elevate the signal and reduce fluctuations within a low-dimensional subspace. A simple method to achieve this is to introduce a negative feedback loop that selectively attenuates activity within the signal subspace. By increasing the intensity of both the signal and the feedback loop in tandem, the network can maintain the signal at a stable level even when the input signal increases, as the feedback cancels out the signal. Simultaneously, the feedback loop also cancels the fluctuations, effectively increasing the signal-to-noise ratio. Studying this system\textquotesingle s overall dynamics and performance required a dynamic mean-field theory that accounts for structured connectivity (represented by the feedback loop), chaotic dynamics, and finite-size fluctuations. The main results of this theory are presented in Figure \ref{fig:fig2}.

When both feedforward and recurrent inputs are increased, they must locally cancel each other out within each neuron to prevent saturation from the net current. This cancelation is a form of local balance akin to the dynamic balance observed between inhibitory and excitatory populations (E-I balance) \citep{Van_Vreeswijk1996-rm}. In the case of E-I balance, strong cancellation results in highly stable mean activity. In our case, the balance between feedforward and recurrent inputs occurs in multiple directions within the neural activity space. This balance can stabilize a high-dimensional signal, provided the signal dimensions are significantly lower than the overall neural activity space \(N\). This network dynamics result in a ``coding subspace,'' where signals can be accurately decoded by a simple linear readout, enhancing the network's computational efficiency and reliability \citep{kadmon2020predictive,Timcheck2022-ch}.

It may seem that synaptic balance appears sufficient to suppress chaotic fluctuations at the readout completely. However, critical limitations exist in the effectiveness of feedback and synaptic balance mechanisms. As John Von Neumann noted, ``Every network or nervous system has a definite time lag between the input signal and the output response'' {[}cite{]}. Neural systems are replete with transmission delays from various sources, including axonal travel time, synaptic dynamics, and membrane resistance. These delays mean a finite gap between when a neuron spikes and when its postsynaptic counterparts receive the signal. The problem is that delayed negative feedback induces oscillations \citep{glass1988}. Consequently, delayed feedback in neural networks results in synchronous oscillations of neural ensembles \citep{kadmon2020predictive,Timcheck2022-ch}, increasing readout error and reducing the information transmission capacity of the network \citep{Chalk2016-zr}.

The trade-off between fluctuations and oscillations leads to an optimal operating point for efficient coding, determined by the balance strength. The frequency of oscillations in this optimal regime depends on the delay, which varies based on its origin. These oscillations can potentially explain the observed peaks in the spectrum of brain activity.

This simple model illustrates the fruitful exchange of ideas between theoretical neuroscience and statistical physics. The challenge of efficient coding in a basic toy model led to the development of a new dynamic mean-field theory, which extends existing theories of dynamic disordered systems. The insights derived from this theory offer valuable perspectives on brain function and may help explain physiological observations. This example is one of many in which the study of neural dynamics and computation in theoretical neuroscience has driven advancements in understanding complex systems, underscoring the significant, ongoing contributions of both disciplines.

\section*{Discussion}\label{discussion}
The emergent dynamics in neural networks exemplify how complex macroscopic behaviors can arise from simple microscopic rules. Through the lens of statistical physics, these studies reveal neural networks as a rich source of phenomena characteristic of complex systems, such as phase transitions, symmetry breaking, and spontaneous order. These features directly represent the physical principles governing all matter, demonstrating that the study of neural computation addresses fundamental questions about the physical world.

Neural networks operate far from equilibrium, process external inputs, and adapt their internal states, making them ideal models for studying non-equilibrium dynamics in physics. This characteristic sets them apart from many traditional physical systems and offers unique insights into out-of-equilibrium phenomena. Techniques developed in theoretical neuroscience to understand information processing, computation, and representation have broad applications beyond neurobiology. They can be applied to analyzing financial markets, studying biological systems, and understanding other large disordered systems, often the focus of applied physics research.

Moreover, high-dimensional chaos in large, disordered neural systems showcases another critical area where neuroscience has enriched physics. Traditional linear stability analysis tools fall short in these regimes, necessitating new methods to understand and control chaos. These methods, crucial for neuroscience, also enhance our ability to manage chaotic behavior in various physical and engineering systems out of equilibrium with quench disorder, potentially contributing to other fields.

The study of neural networks has also led to advancements in our understanding of critical phenomena and universality in non-equilibrium systems. For instance, the concept of criticality at the edge of chaos has provided new perspectives on how biological systems might optimize their information-processing capabilities.

Overall, this essay demonstrates that studying the brain--- a paragon of complexity---necessitates and benefits from theoretical frameworks borrowed from physics. Simultaneously, this research contributes to developing these frameworks, offering profound insights into the behavior of nonequilibrium statistical systems and advancing our understanding of complex systems in general. As we continue to unravel the mysteries of neural computation, we can expect further cross-pollination of ideas between neuroscience and physics, potentially leading to breakthroughs in both fields and opening new avenues for understanding and harnessing complexity in natural and artificial systems.

\section*{Declarations}
\subsection*{Ethical approval}
Not Applicable

\subsection*{Funding}
The author is funded by the Azrieli Foundation. The Research presented was funded by the Gatsby Charitable Foundation and the Swartz Foundation.


\bibliographystyle{spbasic}
\bibliography{paperpile,references}

\end{document}